\documentclass[pdflatex,sn-nature]{sn-jnl}

\usepackage{graphicx}%
\usepackage{multirow}%
\usepackage{amsmath,amssymb,amsfonts}%
\usepackage{amsthm}%
\usepackage{mathrsfs}%
\usepackage[title]{appendix}%
\usepackage{xcolor}%
\usepackage{textcomp}%
\usepackage{manyfoot}%
\usepackage{booktabs}%
\usepackage{algorithm}%
\usepackage{algorithmicx}%
\usepackage{algpseudocode}%
\usepackage{listings}%

\begin{document}

\title[Article Title]{Spatiotemporal toroidal light beams with arbitrary polarization and orientation through a multimode fiber}

\author*[1]{\fnm{Andrew V.} \sur{Komonen}}\email{a.komonen@uq.edu.au}
\author[2]{\fnm{Nicolas K.} \sur{Fontaine}}
\author[1]{\fnm{Martin} \sur{Pl\"oschner}}
\author[1]{\fnm{Marcos Maestre} \sur{Morote}}
\author[2]{\fnm{David T.} \sur{Neilson}}
\author[1]{\fnm{Joel} \sur{Carpenter}}
\author*[1]{\fnm{Mickael} \sur{Mounaix}}\email{m.mounaix@uq.edu.au}

\affil[1]{\orgdiv{School of Electrical Engineering and Computer Science, The University of Queensland, Brisbane, QLD, 4072, Australia}}

\affil[2]{\orgdiv{Nokia Bell Labs, 600 Mountain Avenue, Murray Hill, NJ 07974, USA}}

\abstract{Optical toroidal beams, with donut-shaped intensity profiles and orbital angular momentum (OAM), are promising for applications such as optical manipulation, metrology, and advanced light-matter interactions. However, practical implementations are limited by challenges in controlling their full 3D geometry and the orientation of their OAM.  In this paper, we experimentally demonstrate high-dimensional, polarization-resolved, programmable 3D spatiotemporal toroidal beams with arbitrary 3D geometry. The beams are delivered after propagation through an optical multimode fiber (MMF) that supports 90 spatial/polarization modes. However, if desired, this system can also deliver these beams directly into free space as well. Our approach leverages 25,000 programmable spatiotemporal and polarization degrees of freedom to achieve precise manipulation of the amplitude, phase, polarization and temporal properties of toroidal beams. These beams feature highly customizable 3D geometries, allowing independent control of their aspect ratio and orientation. We further demonstrate the generation of beams with arbitrary OAM orientation, with beam rotations about any 3D spatiotemporal axis. These beams are delivered through an MMF which enables their transport deep into scattering materials and into otherwise hard-to-access regions which could include biological tissues. Hence, this device could enable the application of completely customizable optical manipulations, including rotations, deep within these materials.}

\maketitle

\section{Introduction}

Spatiotemporal light beams, which combine precise control of light’s spatial structure and temporal evolution, are opening new avenues in optics and photonics~\cite{roadmap_spatiotemp_light}. By shaping how light behaves in space and time~\cite{mounaix_time_2020, cruz-delgado_synthesis_2022}, reconfigurable spatiotemporal light fields are exploring new physical phenomena and advancing technologies within ultrafast metrology~\cite{cheng_metrology_2025}, exciting new nonlinear effects~\cite{vieira_nonlinear_2014}, generating exotic topological structures~\cite{forbes_structured_2021} and high capacity optical communication~\cite{wang_optical_comms}.


Toroidal beams of light represent a distinct and compelling subclass of spatiotemporal light fields. These three dimensional beams, defined by their donut-shaped intensity profile, possess the ability to carry orbital angular momentum (OAM) through a spirally varying phase around the toroid~\cite{wan_review, zhan_spatiotemporal_tutorial}. For example, phase wrapping around the tube of the toroid - poloidal direction - creates internally circulating OAM ~\cite{allen_orbital_1992, speirits_waves_2013, wan_toroidal_2022}. Phase wrapping around the singularity of the toroid - toroidal direction - creates OAM that is oriented along the central axis passing through the toroidal hole. First realized experimentally in free space~\cite{wan_toroidal_2022,zdagkas_observation_2022}, toroidal beams combine spatiotemporal beam shaping~\cite{roadmap_multimode,roadmap_wavefront_shaping} with the physics of OAM. This makes them uniquely suited for a multitude of applications, such as enabling precise optical manipulation to rotate, trap, or sort particles~\cite{stilgoe_controlled_2022}. Furthermore, toroidal beams facilitate spin-orbit coupling effects, important for optical manipulation, where OAM interacts with spin angular momentum (SAM) to produce targeted spin-orbit coupling phenomena~\cite{chen_spinorbit_2022}.

The structured intensity and phase profiles of toroidal beams also make them ideal candidates for high-capacity optical telecommunications, where they can encode data into their spatial modes for efficient and robust signal transmission~\cite{wang_optical_comms, huang_spatiotemporal_2024_OAM_comms, shen_nondiffracting_2024, wang_terabit_2012}. Also by tuning their OAM, superluminal and subluminal pulse propagation can be controlled, enabling precision timing and advanced signaling while avoiding the distortions caused by medium dispersion~\cite{mazanov_fast_and_slow}. Beyond telecommunications, their inherent chirality has found applications in molecular biology and nanotechnology, enabling the detection, trapping~\cite{Chiral_trap}, and manipulation of chiral particles~\cite{mun_chiral_detect, ni_gigantic_OAM_detect, ni_giant_OAM_detect}. In quantum optics, toroidal beams are opening new frontiers, allowing the generation of high-dimensional quantum states, advancing quantum communication protocols, and providing a platform for testing emerging quantum field theories~\cite{yang_non-classical_2021}. Hence, toroidal beams represent a valuable tool for the photonics community.

Currently, given the complexity of generating 3D toroidal beams, most kinds of these beams are generated in 2D as spatiotemporal optical vortices (STOV)~\cite{jhajj_spatiotemporal_2016_first, hancock_free-space_2019, piccardo_broadband_2023} including with transverse OAM~\cite{chong_generation_2020_transverse, adams_spatiotemporal_2024}. However, to fully realize the potential applications of toroidal beams, these capabilities need to be extended by controlling two critical aspects of their structure: their spatiotemporal geometry; and the orientation (or tilt) of their OAM. Arbitrary OAM tilt is essential to enable complete 3D optical manipulation as the OAM could be applied as an optical spanner about any 3D axis~\cite{wan_photonic_2022_two_OAM_beam}. Arbitrary OAM tilt could also enable other applications including enhanced optical tweezing control~\cite{banzer_photonic_2013, aiello_transverse_2015},  and enable specific beam tailoring for complex light-matter interactions ~\cite{sederberg_vectorized_2020, forbes_sculpting_2020}. However, achieving arbitrary 3D control over the tilt of the OAM remains a challenge. For instance, existing techniques, which have arbitrary 3D tilt control include the numerically simulated use of photonic crystal slabs with physical tilts~\cite{wang_engineering_2021_photonic_crystal}, experimentally demonstrated combination of two OAM beams~\cite{wan_photonic_2022_two_OAM_beam}, or employing experimental setups, without arbitrary spatiotemporal control, that require manually tilting lenses which degrades beam quality beyond 60 degrees~\cite{zang_spatiotemporal_2022_tilt_lens}. Therefore, it has not previously been possible to generate a high fidelity single 3D OAM toroidal beam in free space at any arbitrary orientation. Furthermore, these approaches are typically restricted to generating circular vortex configurations, lacking the ability to create arbitrarily defined 3D toroidal structures. Beyond geometric constraints, full control over the amplitude, phase~\cite{wan_scalar_2022_hopfions, zhong_toroidal_2024_phase_toroidal}, and polarization~\cite{fang_vectorial_2021} of toroidal beams has not yet been achieved. Unlocking such complete control across all degrees of freedom of toroidal beams would pave the way for new and complex applications, including dynamically tunable optical tweezers for precise particle manipulation~\cite{roadmap_optical_tweezers} and as a metrological tool to analyze complete 3D motion across all light's degrees of freedom~\cite{cheng_metrology_2025}.

Here, we demonstrate complete experimental control over high-dimensional, polarization-resolved, arbitrarily oriented 3D OAM toroidal beams. This is achieved with precise control of their amplitude, phase, and polarization, after propagation through a graded-index multimode fiber (MMF) supporting 90 spatial/polarization modes. By "high-dimensional" we refer to the ability to manipulate the linear superposition of 45 Hermite-Gaussian modes per polarization state. This capability allows us to define the aspect ratio of the toroidal beam, characterized by $\frac{R}{r}$ where $R$ is the major radius (the distance from the toroid center to the tube center) and $r$ is the minor radius (the radius of the tube). Such control enables the center hole size to vary throughout the toroid by the choice of $R$ and $r$. Unlike a ring torus or Laguerre-Gaussian modes, which in 3D exhibit an approximately cylindrical singularity cross section, our approach allows for arbitrary deformation of the toroid to create completely configurable complex structures such as the toroidal singularity in 3D. Our setup provides a solution to generate and deliver custom toroidal beams, controlled across all degrees of freedom, to any challenging-to-access location, paving the way for the development of new applications in ultrafast optics and photonics.

\section{Results}
\subsection{Definition of the toroidal beam}

\begin{figure} [h]
  \centering
  \includegraphics[width=\textwidth]{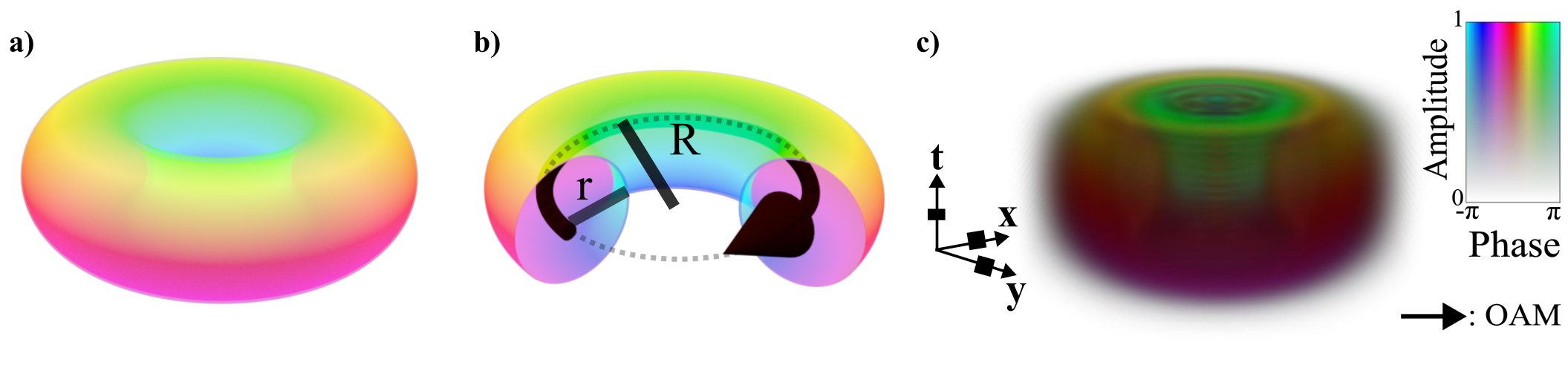}
  \caption{a) Target torus as defined in Eq.~\ref{eq:torus}. The minor radius ($r$) and major radius ($R$) set the complete 3D structure. The beam features a spiral phase varying along the poloidal direction with a topological charge 1. (b) Such phase profile creates an internally circulating OAM. (c) The torus is now sliced along the temporal axis in 20 discrete 2D cross sections with an experimental temporal resolution of around 226fs. Each slice is then overlapped onto a basis made of 45 Hermite-Gaussian modes. The full structure forms the goal beam our apparatus aims to generate.  Axis ticks: $x$ and $y$: 100$\mu$m, $t$: 200 fs}
  \label{fig:Simulated Donut}
\end{figure}

The beams we aim to experimentally generate are optical toroidal beams. A toroidal beam is a three-dimensional structure: two transverse dimensions ($x,y$) and one longitudinal dimension ($t$). 
The toroidal beam is defined by the following set of parametric equations~\cite{moroni_toric_2017_toroid_maths}:

\begin{align}
    x &= (R + r \cos v) \cos u \nonumber\\
    y &= (R + r \cos v) \sin u \label{eq:torus} \\
    t &= r \sin v \nonumber
\end{align}

$R$ and $r$ represent respectively the major and the minor radius of the torus. $u$ and $v$ are two parameters that range from 0 to $2\pi$, representing respectively the toroidal (around the large circular axis) and the poloidal (around the torus cross-section) angle. The non-intensity zone of the torus is hence defined along the $t$-axis, which corresponds in the experiment to the time/delay axis. 

Here we assume a discrete definition of the optical toroidal beam along the longitudinal (time) dimension. Hence, we can define the optical toroidal beam as discrete 2D transverse cross sections at different delays. By encoding phase to each individual 2D section, the desired spatiotemporal field can be precisely tailored. The derived calculation from Eq.~\ref{eq:torus} to define the torus as well as defining its phase is presented in the Methods. The toroidal beam can also be defined and generated in orthogonal polarization states being in horizontal ($H$) or vertical ($V$) or both states simultaneously.

Using Eq.\ref{eq:torus}, we simulated the 2D spatial cross sections required to create a 3D spatiotemporal toroidal beam, as illustrated in Fig.\ref{fig:Simulated Donut}. In this example, the toroidal beam is encoded with a poloidal spiral phase to generate an internally circulating OAM (arrow in Fig.~\ref{fig:Simulated Donut}b), a characteristic feature of toroidal beams~\cite{wan_toroidal_2022}. 

\subsection{Experimental setup}

Generating such spatiotemporal toroidal optical beams experimentally requires full independent control over both the spectral and  transverse properties of the beam. While most previous approaches focus on manipulating only two of the three required dimensions, we recently developed an apparatus capable of generating arbitrary vector spatiotemporal beams~\cite{mounaix_time_2020}. A simplified setup is presented in Fig.~\ref{fig:Setup}. The full apparatus is described in  Supplementary Fig.~1. 
A swept-wavelength laser source is coupled into a polarization-resolved programmable wavelength selective switch (WSS)~\cite{neilson_wavelength_2006}. Within the WSS, a liquid crystal on silicon spatial light modulator (SLM) is located in the Fourier plane of a diffraction grating. Polarization diversity optics within the WSS enables imaging of the two orthogonal polarization states side-by-side on the SLM.

\begin{figure} [h]
  \centering
  \includegraphics[width=\textwidth]{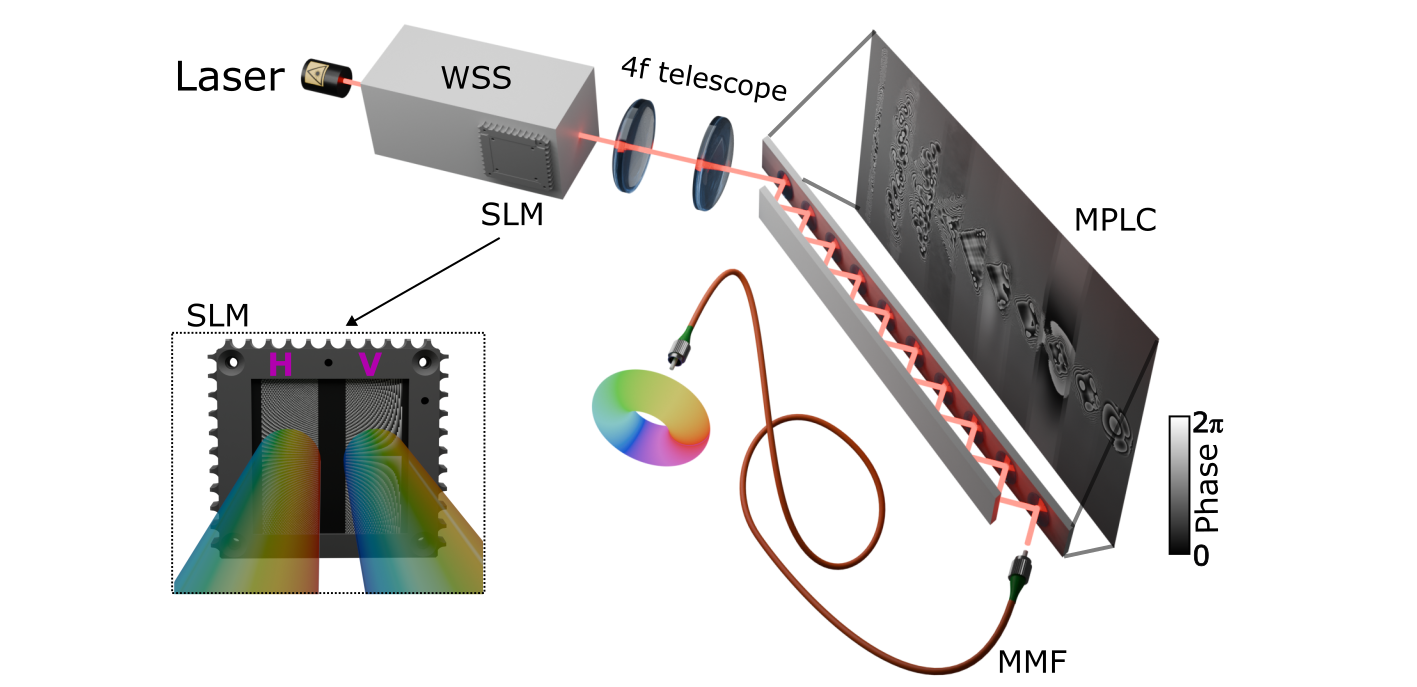}
  \caption{Simplified apparatus enabling the generation of 3D toroidal beams. The system combines a wavelength-selective switch (WSS) with a multi-plane light conversion (MPLC) device. Inset SLM: zoomed view of the spatial light modulator (SLM) within the WSS, featuring a computer-generated hologram with distinct patterns on the left and right sides to control the two polarization states independently. The horizontal axis of the hologram encodes different wavelengths, for illustration in different colors. Inset MPLC: Lithographically etched holograms of the MPLC, which is a 45 Hermite-Gaussian mode-sorter with 10 phase masks. MMF: multimode fiber.}
  \label{fig:Setup}
\end{figure}

The SLM displays a programmable hologram that can shape the spatiotemporal field for each polarization state. The spectral/temporal properties of the beam are controlled along the horizontal axis of the SLM, with a spectral resolution of 15 GHz and a total controllable spectral bandwidth of 4.4 THz (35 nm), centered at 193 THz (1551.6 nm). In the temporal domain, this corresponds to a resolution of around 226 fs. Each wavelength can also be independently manipulated along the vertical axis of the SLM, enabling a customizable vertical spot array to be generated at the output of the WSS. 


After the WSS, this spot array is directed to a multi plane light conversion (MPLC) device~\cite{fontaine_laguerre-gaussian_2019}. The MPLC, consisting here of 10 lithographically etched phase masks in cascade and a flat mirror, maps these vertically dispersed spots into a set of 45 orthogonal Hermite-Gaussian modes. This process transforms the 2D beam from the WSS into a fully 3D beam~\cite{mounaix_time_2020}. The beam is then imaged on a 5-meter-long graded-index MMF with 50 $\mu$m core, supporting 90 spatial/polarization modes. The generated beam is located at the output of the MMF. In Fig.~\ref{fig:Setup}, we represent a toroidal beam with a $2\pi$ phase wrap along the toroidal axis. To ensure we generated the targeted beam, we measure the polarization-resolved spectrally-resolved field distribution,  via swept-wavelength off-axis digital holography\cite{mounaix_control_2019_TM,carpenter_digholo_2022}. 

Due to modal dispersion and mode coupling within the MMF, the spatiotemporal field differs significantly from the input field. To fully characterize this linear transformation, we measure its transmission matrix (TM) $\textbf{T}(\lambda)$\cite{popoff_measuring_2010_TM} with spectral resolution, which linearly maps the measured spectrally-resolved field at the output of the MMF to the input field launched with the SLM. Once the complete spectrally-resolved transmission matrix of the system is measured, it becomes possible to experimentally generate toroidal beams at the output of the MMF.

Our apparatus enables the control of more than 25,000 independent spatiotemporal degrees of freedom of the input light beam: 90 spatial/polarization modes on the vertical axis of the SLM, and around 290 spectral components on the horizontal axis of the SLM. To achieve a 3D target toroidal beam at the output of the MMF, the required field to be launched from the SLM is 
calculated using the conjugate transpose operator of the spectrally-resolved TM $\textbf{T}(\lambda)$~\cite{mounaix_spatiotemporal_2016}. This input field compensates for the spatiotemporal distortions induced by the propagation within the MMF to achieve the desired output (see Methods). 
To experimentally generate this input field, a computer generated hologram is calculated through a modified Gerchberg-Saxton (GS) algorithm~\cite{mounaix_time_2020}, and then displayed on the SLM (see Methods). Once the hologram is displayed on the SLM, swept-wavelength off-axis digital holography is performed to measure the spatiotemporal field at the MMF output in the spectral domain. The corresponding spatiotemporal field in the temporal domain is then obtained by applying a Fourier transform along the spectral axis.

\subsection{Experimental results}

\begin{figure} [h]
  \centering
  \includegraphics[width=\textwidth]{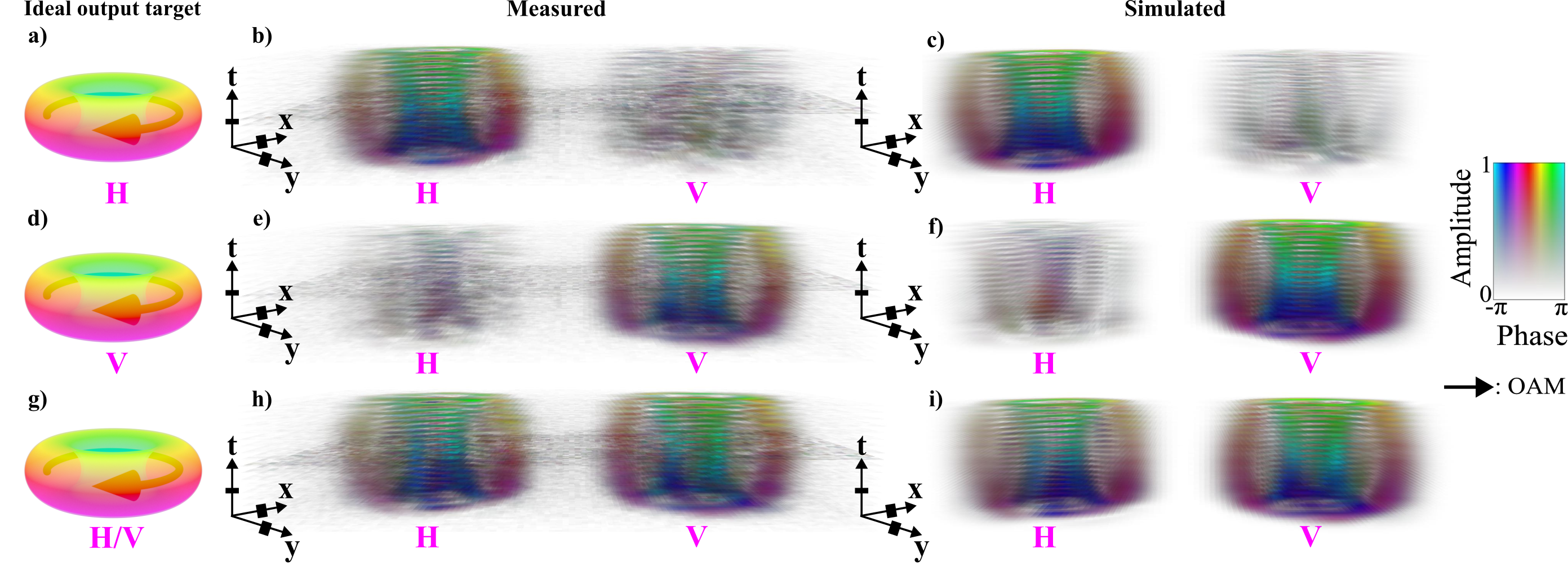}
  \caption{Ideal target, experimentally generated and numerically simulated 3D spatiotemporal toroidal beams with a duration of $4.5$ ps for a-c) horizontal polarization, d-f) vertical polarization, and g-i) simultaneous horizontal and vertical polarizations. All toroidal beams exhibit a total phase wrapping of $2\pi$ along the poloidal direction, corresponding to a topological charge of 1, which generates internally circulating OAM. The simulated beams represent the numerically propagated fields of the input beam displayed on the SLM. Axis ticks: $x$ and $y$: 100$\mu$m, $t$: 200 fs}
  \label{fig:Toroidal Beams}
\end{figure}

A set of experimentally measured 3D toroidal optical beams at the output of the MMF is presented in Fig.~\ref{fig:Toroidal Beams}. We demonstrate a toroidal beam defined over a delay span of $4.5$~ps,  corresponding to 20 cross sections along the temporal axis. Note that here we are plotting each individual 3D beam as a sequence of these 2D cross sections so that their internal complex fields can be seen. Fig.~\ref{fig:Toroidal Beams} also highlights the polarization control achieved in the generated beams: we show  horizontally polarized, vertically polarized, or 45-degree polarized toroidal beams. Each toroidal beam is encoded with a poloidal phase wrap of topological charge 1, resulting in internally circulating OAM. 

To evaluate the quality of the experimentally measured fields, that we name $\vec{E}^{\text{out,measured}}$, we compare them with the ideal target field the system can generate. Using the spectrally-resolved transmission matrix, the expected output field $\vec{E}^{\text{out,expected}}$ can be numerically calculated (see Methods). $\vec{E}^{\text{out,expected}}$ is shown in Fig.~\ref{fig:Toroidal Beams} as "Simulated".

To assess the experimental results, we calculate the overlap integral $\mathcal{O}$ between the experimentally measured field $\vec{E}^{\text{out,measured}}$ and the numerically propagated field $\vec{E}^{\text{out,expected}}$ (See Methods). The overlap integral $\mathcal{O}$ was calculated for the three beams presented in Fig.~\ref{fig:Toroidal Beams}, the horizontally, vertically, and 45-degree polarized beams. The magnitude squared of these overlaps were $|\mathcal{O}|^{2}=80\%$, $|\mathcal{O}|^{2}=82\%$ and $|\mathcal{O}|^{2}=81\%$ respectively. These results highlight the system’s strong phase control, as even minor experimental deviations in phase, or slight change in the TM, would significantly affect the overlap, demonstrating the high fidelity of the generated beams.

\begin{figure} 
  \centering
  \includegraphics[width=0.6\textwidth]{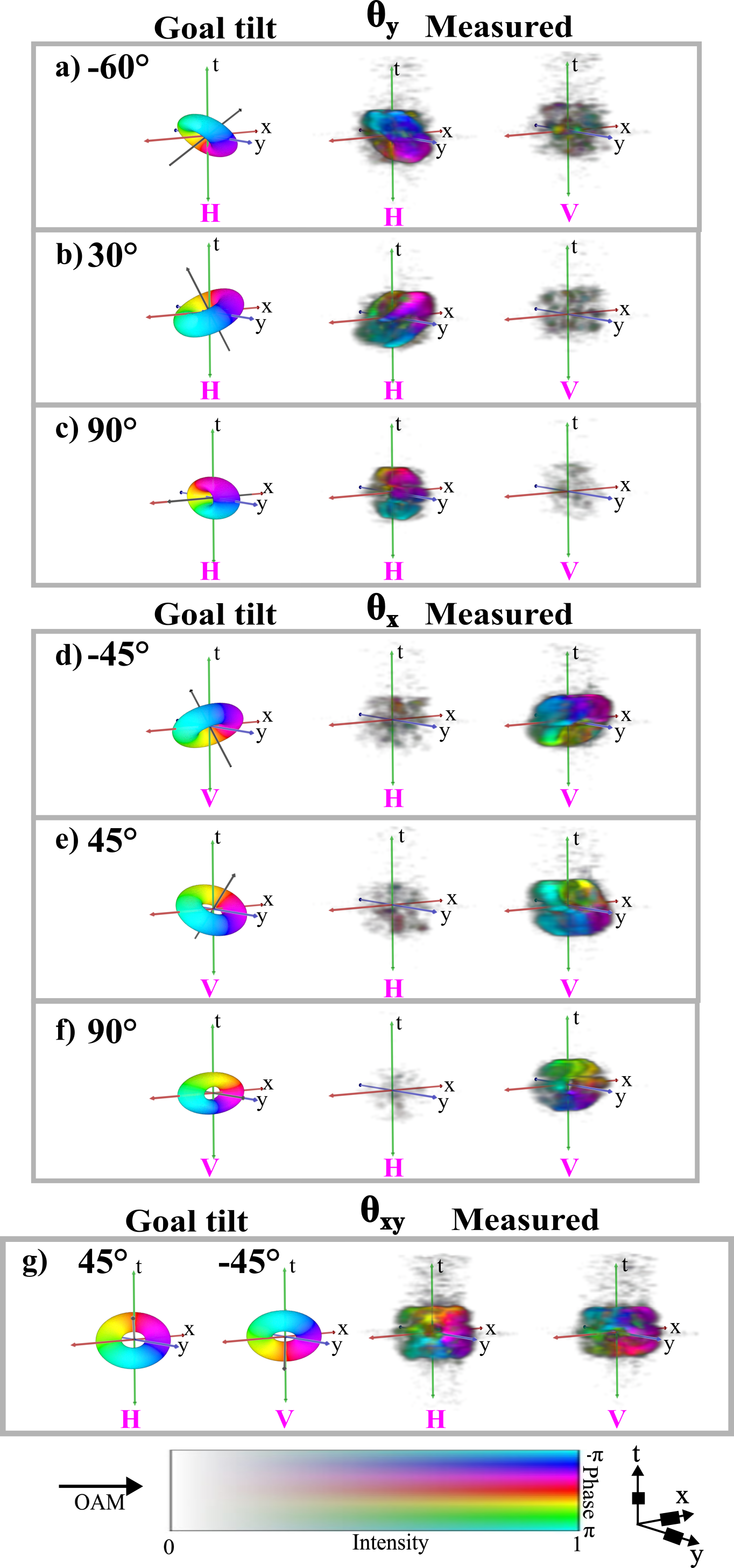}
  \caption{Ideal target and experimentally generated 3D spatiotemporal toroidal beams at arbitrary orientation angles. The beams are rotated about: a)-c) the $y$-axis, with angles of $-60^\circ$, $30^\circ$, and $90^\circ$, shown for horizontal polarization; d)-f) the $x$-axis, with angles of $-45^\circ$, $45^\circ$, and $90^\circ$, shown for vertical polarization; and g) off-axis orientations, where the toroidal beam is rotated by $45^\circ$ for both the $x$- and $y$-axes for horizontal polarization, and $-45^\circ$ for both axes for vertical polarization. These toroidal beams have a duration of $4.5$ ps. All toroidal beams exhibit a total phase wrapping of $2\pi$ along the toroidal direction, corresponding to a topological charge of 1, which generates an orbital angular momentum (OAM) vector passing through the toroidal singularity. The complete field animations for a)-g) are presented in Supplementary Video 1-7. Axis ticks: $x$ and $y$: 500$\mu$m, $t$: 1ps
}
  \label{fig:Toroidal Beams at arbitary orientation}
\end{figure}

We further extend our capacity to generate 3D toroidal optical beams to arbitrary geometric orientations of the singularity, as shown in Fig.~\ref{fig:Toroidal Beams at arbitary orientation}. This was achieved by analytically tilting the torus (See Methods). In Fig.~\ref{fig:Toroidal Beams at arbitary orientation}, we demonstrated 3D toroidal beams defined over a delay span of $4.5 $ ps with various orientations and various output polarization states, highlighting the versatility and applicability of our approach.

In this configuration, the topological charge remains set to 1, but the phase wrapping is now defined along the toroidal direction, causing the phase to wrap around the singularity. This toroidal phase wrapping generates OAM aligned with the tilt angle, and consequently with the axis of the singularity. Therefore, this not only further highlights the device's versatility to enable arbitrary phase encoding. It also is highly desired for practical applications, such as optical manipulation, as it allows for arbitrarily directed OAM in 3D. To show these possibilities, in Fig.~\ref{fig:Toroidal Beams at arbitary orientation}a-c), we show toroidal beams rotated about the $y$-axis with squared magnitude overlaps of $|\mathcal{O}|^2 = 82\%$, $|\mathcal{O}|^2 = 81\%$ and $|\mathcal{O}|^2 = 79\%$ respectively. Therefore, here we have obtained 2D spatiotemporal control of these OAM beams showing that we can arbitrarily rotate them around the y axis with high beam fidelity. Next, in Fig.~\ref{fig:Toroidal Beams at arbitary orientation}d)-f), we show toroidal beams which are rotated about the $x$-axis with squared magnitude overlaps of $|\mathcal{O}|^2 = 83\%$, $|\mathcal{O}|^2 = 81\%$ and $|\mathcal{O}|^2 = 82\%$ respectively. Therefore, by showing these beams rotated around the $x$ and $y$ axes with high beam fidelity, we have full 3D spatiotemporal control. This allows us to generate a toroidal beam with OAM aligned to any spatiotemporal axis in 3D. To directly demonstrate this we see, in Fig.~\ref{fig:Toroidal Beams at arbitary orientation}g) the toroidal beam is rotated about an arbitrary 3D axis with squared magnitude overlap of $|\mathcal{O}|^2 = 82\%$. Therefore, Fig.~\ref{fig:Toroidal Beams at arbitary orientation}g) highlights both the flexibility and practicality of our approach. It demonstrates that we are not limited to rotating the OAM direction along a specific transverse axis ($x$ or $y$). By simply reprogramming the SLM hologram, without physically changing the experiment or moving any detector or receiver, the OAM direction could be adjusted freely in 3D. This could allow, for example, for dynamically varying optical torques to be applied in 3D spatiotemporal space in order to sustain an optical trap. Or even performing optical manipulations to dynamically control and adjust the movement of a sample in any direction. Fig.~\ref{fig:Toroidal Beams at arbitary orientation}g) also shows that we have a toroidal beam where the OAM singularity is oriented differently for its horizontal and vertical polarization states. For the horizontally polarized beam, the OAM singularity is rotated at $+45^\circ$ relative to both the $x$- and $y$-axes (i.e., $+45^\circ$ on $x$ and $+45^\circ$ on $y$). For the vertically polarized beam, the OAM singularity is rotated at $-45^\circ$ relative to both axes (i.e., $-45^\circ$ on $x$ and $-45^\circ$ on $y$). This polarization control highlights that this system is directly applicable to polarization sensitive applications such as those requiring circular polarization for example for chirality detection and manipulation. 

Utilizing both the device's polarization control to generate SAM along with the arbitrary directional control of OAM could enable interesting spin-orbit coupling investigations. This is since SAM and OAM will both act independently on a given target and could also be used in specific cases to detect targets with a given sensitivity to either SAM or OAM. By utilizing this spin-orbit coupling control, these toroidal beams could be used to enhance complex structures such as toroidal knots and links~\cite{dennis_isolated_2010, kedia_tying_2013, zhan_spatiotemporal_tutorial}. This is since when generating these beams, the interaction of SAM and OAM directly influences both their topological structure and physical properties such as polarization~\cite{sugic_singular_2018}.

Therefore, these results highlight the system’s ability to generate toroidal beams at arbitrary orientations, beyond simple on-axis configurations. With this capability, we can rotate toroidal beams around any 3D spatiotemporal axis, allowing for OAM generation on any 3D spatiotemporal axis, with full control over their polarization, phase wrapping, and orientation angle. Applying these capabilities will enhance the creation of complex structures, such as toroidal knots and links, whose topological and physical properties can be completely controlled.

\section{Discussion}

We have demonstrated the experimental generation of arbitrary polarization-resolved 3D spatiotemporal toroidal beams, with the potential to extend this approach to other exotic spatiotemporal beams types such as optical hopfions. Optical hopfions have particle-like structures which could enable: novel excitations in matter and nanostructures, novel metrological applications and high-dimensional optical communications \cite{wan_scalar_2022_hopfions, zhong_toroidal_2024_phase_toroidal}. Utilizing this system, 3D spatiotemporal optical hopfions could be generated with arbitrary orientations and control over all degrees of freedom. This would follow the same procedure presented in this paper, with only the phase encoding needing to be changed. 

Our ability to deliver toroidal beams to hard-to-access locations mitigates the limitations imposed by current methods to create STOVs, which rely on bulky free-space components, hence limiting the flexibility and practicality of such beams in many applications~\cite{stellinga_time--flight_2021}. Generating toroidal beams directly at the output of a MMF, which can be easily routed to deliver the beam to the desired location is a promising solution. MMFs provide significant advantages through their ability to access environments where light delivery is challenging due to absorption or multiple scattering~\cite{leite_three-dimensional_2018}, making them ideal for applications in telecommunications~\cite{matthes_learning_2021}, fiber power amplification~\cite{rothe_output_2024,zervas_high_2014}, high photon-efficiency systems~\cite{cao_controlling_2023_MMF}, optical tweezers~\cite{cao_controlling_2023_MMF} and for nonlinear interactions~\cite{wright_physics_2022}. For instance, MMFs enable optical forces to be applied deep within biological tissues, extending the reach of non-invasive, non-destructive optical manipulation and imaging~\cite{cao_controlling_2023_MMF, ploschner_compact_2015_MMF}. Our apparatus not only enables compensation of mode coupling and modal dispersion, it can also overcome other complex effects inherent to MMFs~\cite{finkelstein_spectral_2023}. Thus, by utilizing an MMF, the practicality of these toroidal beams is enhanced, and the door is opened to novel applications in previously inaccessible environments.


\section{Methods}\label{sec11}

\subsection{Defining the torus and applying phase}

\subsubsection{Defining the torus}

The set of parametric equations defining the torus (Eq.~\ref{eq:torus}) satisfy the following equation:

\begin{equation}
    \Big(\sqrt{x^2 + y^2}-R\Big)^2 + t^2 - r^2 = 0
\label{eq:torus_solve}
\end{equation}

Solving Eq.~\ref{eq:torus_solve} provides the coordinates of the points on the surface of the torus. 

\subsubsection{Defining the phase of the torus}

The toroidal phase $u$ can be retrieved from the parametric equations defining the torus (Eq.~\ref{eq:torus}):

\begin{equation}
    u=\tan\frac{y}{x}
\label{eq:toroidal_phase}
\end{equation}

Similarly, the poloidal phase $v$ reads:

\begin{equation}
    v=\tan\frac{t}{\sqrt{x^2+y^2}-R}
\label{eq:poloidal_phase}
\end{equation}

A toroidal phase wrap is achieved if  $\varphi=u$, with $\varphi$ the phase of the torus. Similarly, a poloidal phase wrap is achieved if  $\varphi=v$.

\subsubsection{Defining the torus with a tilt}
Rotating the torus around one axis is achievable simply by using the corresponding rotation matrix. For example, let's define the toroidal beam from Fig.~\ref{fig:Toroidal Beams at arbitary orientation}d, with a tilt $\theta$ along the $x$-axis. The rotated $y'$ and $t'$ reads:

\begin{align}
y'&= y\cos\theta + t\sin\theta \nonumber \\ 
t'&= -y\sin\theta + t\cos\theta 
\label{eq:rotation_matrix_x}
\end{align}

Substituting Eq.~\ref{eq:rotation_matrix_x} into Eq.\ref{eq:torus}, the following set of equations defines the tilted torus as a function of $\theta$:

\begin{align}
    x &= (R + r \cos v) \cos u \nonumber\\
    y' &= (R + r \cos v) \sin u \cos\theta + r \sin v \sin\theta\label{eq:torus_tilt_thetax} \\
    t' &= -(R + r \cos v) \sin u \sin\theta + r \sin v \cos\theta \nonumber
\end{align}

Finding the points on the surface of the torus is achieved by solving the following equation:

\begin{equation}
    \Big(\sqrt{x^2 + y'^2}-R\Big)^2 + t'^2 - r^2 = 0
\label{eq:torus_solve_tilt}
\end{equation}

The above method can be applied for a tilt along the $y$-axis, by changing Eq.~\ref{eq:rotation_matrix_x} with the corresponding rotation matrix.

\subsection{Producing tilted toroidal beam figure}

In Fig.~\ref{fig:Toroidal Beams}, we have plotted the amplitude of the 3D beams with linear transparency including all measured points. In this figure, we also plot using a sequence of 2D cross sections, which was done for increased clarity to see the internal component of the poloidally varying phase. It also shows that we can control the internal structure which surrounds the center hole of these 3D toroidal beams.

In Fig.~\ref{fig:Toroidal Beams at arbitary orientation}, for improved clarity of the results, we plotted using a volumetric plot so that we can clearly see each beams' 3D structure including its tilt angle. For each tilted angle, we have plotted the intensity (amplitude squared) of the measured data, whilst visually highlighting the surface of the toroid by linearly shifting the RGB scale from 0 - 255 to 0 - $\frac{255}{64}$. Also to improve clarity a quadratic transparency profile was used with points containing the lowest 7\% of power across both polarizations also being removed.

\subsection{Measuring the spectrally-resolved transmission matrix of the system}

The process involves sequentially launching each of the 45 modes per input polarization state across the entire spectral bandwidth, achieved by programming appropriately the SLM within the WSS to generate the desired input modes. Once the hologram is displayed, the laser source is swept through the spectral range, and the spectrally-resolved polarization-resolved fields are measured on the camera using digital holography. The camera is electronically triggered during each sweep to capture 362 frames evenly spaced in frequency across the spectral bandwidth, slightly oversampling over the spectral axis. Each polarization-resolved measured field is overlapped onto a set of 45 orthogonal Hermite-Gaussian modes $\vec{E}^{out}$ to analyze the mode decomposition, resulting in a $362\times90$ matrix that captures the mode coupling and modal dispersion for the specific input mode launched $\vec{E}^{in}$. This process is repeated for all input modes across both polarization states, resulting in a complete TM with dimensions $362\times90\times90$ defined by:

\begin{equation}
\vec{E}^{out}(\lambda) = \textbf{T}(\lambda) \vec{E}^{in}(\lambda)
\label{eq:TM}
\end{equation}

\subsection{Defining the input field to achieve a target field at the output of the MMF}

The 3D target toroidal beam $\vec{E}^{\text{out,target}}$ is defined by solving Eq.~\ref{eq:torus_solve}. The beam is then sliced into 2D cross sections. Each polarization-resolved transverse cross section is decomposed into an orthogonal set of 45 Hermite-Gaussian modes. This decomposition yields 45 coefficients, each with amplitude and phase, encapsulating the mode composition of each cross section. A 1D Fourier transform is then applied along the temporal axis to represent the 3D target toroidal beam in the spectral domain, as a superposition of 45 Hermite-Gaussian modes per polarization state for each wavelength. The resulting target vector thus has a dimension of $362\times90$.
To generate such a beam at the output of the MMF, the required input field is calculated using the conjugate transpose operator of the spectrally-resolved TM $\textbf{T}(\lambda)$~\cite{mounaix_spatiotemporal_2016}. This calculation provides the input field $\vec{E}^{\text{inject}}$, that must be launched into the MMF to achieve the desired target toroidal beam at the output.

\begin{equation}
\vec{E}^{\text{inject}}(\lambda) = \textbf{T}^\dagger(\lambda) \vec{E}^{\text{out,target}}(\lambda)
\label{eq:TM_invert}
\end{equation}

The input field $\vec{E}^{\text{inject}}$ is a spatiotemporal field defined as a superposition of 45 Hermite-Gaussian modes per polarization state for each spectral component. 

\subsection{Phase mask calculation for generating optical toroidal beams}

Experimentally, $\vec{E}^{\text{inject}}$ is generated using a computer generated hologram displayed on the SLM within the WSS.

The system is first calibrated through the programmable SLM to determine the phase profile required along its vertical direction to selectively excite each of the 45 individual modes supported by the MPLC for every wavelength and each polarization state. Each transverse mode of the MPLC corresponds then to a specific physical position along the 1D vertical axis in the Fourier plane of the SLM. 

Building on this calibration, the generation of 3D spatiotemporal toroidal beams involves programming a custom phase mask onto the SLM. This phase mask creates a spot array of 45 distinct spots, where the light directed to each spot is precisely controlled in both amplitude and phase, for every addressable wavelength component and polarization state. By mapping these spots to the corresponding modes in the MPLC, the system synthesizes the desired toroidal beam with complete control over its spatial, spectral/temporal, and polarization properties, by only manipulating two dimensions on the SLM: the spatial frequency $k_x$ and the spectral frequency $f=c/\lambda$.

To calculate the required phase mask, we use a modified Gerchberg-Saxton (GS) algorithm. The process starts with the target field $\vec{E}^{\text{inject}}(t)$, obtained by applying a Fourier transform to $\vec{E}^{\text{inject}}(\lambda)$ from Eq.~\ref{eq:TM_invert} along the spectral axis. This field is already decomposed into the 45 Hermite-Gaussian (HG) modes supported by the system for each different delay. This step determines the required amplitude and phase for each spot of the input spot array, which maps each spot to the corresponding mode in the MPLC, resulting in a 2D target field $\vec{E}^{\text{target,GS}}((x,t))$, where $x$ represents the vertical axis in space and $t$ represents delay. A 2D Fourier transform applied to this field provides the field in the SLM plane $\vec{E}^{\text{SLM,GS}}(k_x,f)$, where $k_x$ denotes spatial frequency and $f$ denotes spectral frequency. The amplitude distribution of $\vec{E}^{\text{SLM,GS}}(k_x,f)$ along $k_x$ is fixed by the physical size of the beam across the vertical axis of the SLM and remains constant throughout the GS iterations, while only the phase of $\vec{E}^{\text{SLM,GS}}(k_x,f)$ is updated.

The algorithm proceeds by performing a 2D inverse Fourier transform of the SLM plane $\vec{E}^{\text{SLM,GS}}(k_x,f)$ to generate the numerical target field $\vec{E}^{\text{simulated,GS}}(x,t)$ in the $(x, t)$ domain, which is then overlapped with the desired target field $\vec{E}^{\text{target,GS}}(x,t)$. Because the SLM provides only phase control, while the target requires both amplitude and phase control, some loss is inevitable in achieving the desired transformation. To manage this, we define a scattered field region located outside the target area~\cite{mounaix_time_2020} and assign a small portion of the light to this region while enforcing the amplitude and phase of $\vec{E}^{\text{simulated}}(x,t)$ to match $\vec{E}^{\text{target,GS}}(x,t)$ within the region of interest. A 2D Fourier transform is then applied to return $\vec{E}^{\text{SLM,GS}}(k_x,f)$, and the average phase error between this updated plane and the previous SLM plane is removed. The resulting field becomes the new SLM plane.

If the overlap between $\vec{E}^{\text{simulated,GS}}(x,t)$ and $\vec{E}^{\text{target,GS}}(x,t)$ meets the required beam quality, the phase of $\vec{E}^{\text{SLM,GS}}(k_x,f)$ is used as the final SLM phase mask. If the desired beam quality is not achieved, the process repeats, directing more power into the scattered region. This iterative refinement continues until the desired beam quality is achieved with high fidelity, ensuring precise generation of 3D spatiotemporal toroidal beams with minimal but unavoidable hologram loss~\cite{mounaix_time_2020}.

\subsection{Expected output field calculated by numerically propagating the beam}

The expected output field $\vec{E}^{\text{out,expected}}$ is numerically calculated by propagating numerically the input field $\vec{E}^{\text{inject}}$ through the the spectrally-resolved TM as follows:

\begin{align}
\vec{E}^{\text{out,expected}}(\lambda) &= \textbf{T}(\lambda) \vec{E}^{\text{inject}}(\lambda) \nonumber \\ 
&= \textbf{T}(\lambda) \textbf{T}^\dagger(\lambda) \vec{E}^{\text{out,target}}(\lambda)
\label{eq:num_propagated}
\end{align}

To generate the numerically propagated beams shown in Fig.~\ref{fig:Toroidal Beams}, we use $\vec{E}^{\text{out,expected}}$ as defined in Eq.~\ref{eq:num_propagated}. $\vec{E}^{\text{out,expected}}$ represents the expected field at the output tip of the multimode fiber, expressed as a mode superposition within a set of 45 Hermite-Gaussian modes for each output polarization state and wavelength:

\begin{equation}
\vec{E}^{\text{out,expected}}(\lambda) = 
\begin{bmatrix}
a_0 \\
\vdots \\
a_N
\end{bmatrix}
\label{eq:to_launch_vector}
\end{equation}

Here $a_i$ is a complex coefficient that encapsulates the mode decomposition of $\vec{E}^{\text{out,expected}}(\lambda)$ into the $i$-th Hermite-Gaussian mode $\text{HG}_i$. The parameter $N$ in Eq.~\ref{eq:to_launch_vector}  equals 90, representing the total number of Hermite-Gaussian modes across both output polarization states. 
While $\vec{E}^{\text{out,expected}}(\lambda)$ is structured as a vector, its corresponding field representation is referred to as $\vec{E}^{\text{num. prop.}}(\lambda)$. To reconstruct the field $\vec{E}^{\text{num. prop.}}(\lambda)$, we compute the appropriate linear superposition of the Hermite-Gaussian modes in amplitude and phase:

\begin{equation}
\vec{E}^{\text{num. prop.}}(\lambda) = \sum_i a_i(\lambda) \text{HG}_i(\lambda)
\label{eq:mode_superposition}
\end{equation}

After defining $\vec{E}^{\text{num. prop.}}$ across the entire spectral bandwidth, a Fourier transform is applied along the spectral axis to obtain $\vec{E}^{\text{num. prop.}} (t)$ in the temporal domain. This temporal representation is plotted on the  right-hand side, labelled "Simulated" in Fig.~\ref{fig:Toroidal Beams}.

\subsection{Overlap integral}

Since the toroidal beam is defined in the temporal domain, the overlap between the expected field $\vec{E}^{\text{out,expected}}$ and the measured field $\vec{E}^{\text{out,measured}}$ is calculated using the field in this domain, which is obtained by applying a Fourier transform to the spectrally resolved field along the spectral axis. It reads:

\begin{equation}
\mathcal{O} = 
\frac{\int_{\tau_1}^{\tau_2} \int_x \int_y (\vec{E}^{\text{out,measured})*}(t, x, y) \vec{E}^{\text{out,expected}}(t, x, y) \, dx \, dy \, dt}
{||\vec{E}^{\text{out,measured}}(t, x, y)|| \cdot 
||\vec{E}^{\text{out,expected}}(t, x, y)||} 
\label{eq:overlap_integral}
\end{equation}

with
\begin{align}
||\vec{E}^{\text{out,measured}}(t, x, y)|| &= \sqrt{\int_{t=\tau_1}^{t=\tau_2} \int_x \int_y \left| \vec{E}^{\text{out,measured}}(t, x, y) \right|^2 \, dx \, dy \, dt} \nonumber \\
||\vec{E}^{\text{out,expected}}(t, x, y)|| &= \sqrt{\int_{t=\tau_1}^{t=\tau_2} \int_x \int_y \left| \vec{E}^{\text{out,expected}}(t, x, y) \right|^2 \, dx \, dy \, dt} \nonumber
\label{eq:overlap_integral_norm}
\end{align}

Here, $*$ denotes the complex conjugate operator, and ($\tau_1,\tau_2$) is the delay span over which the toroidal beam is defined.


\backmatter

\section*{Declarations}

\subsection*{Data availability}

The measured data are available from the corresponding authors on reasonable request.

\subsection*{Acknowledgements}

The authors acknowledge Daniel Dahl for designing and producing the 3D printed SLM mount and helpful discussions. The authors also acknowledge Jorge Silva for helpful discussions. We acknowledge the Discovery (DP170101400, DE180100009, DE210100934, FT220100103, FT230100388) program of the Australian Research Council, the Westpac Scholars Trust, and NVIDIA Corporation for the donation of the GPU used for this research. 

\subsection*{Author contributions}

Experiments performed by A.V.K. with assistance from M.M.M., and M.P. Optical system designed by N.F., D.N., J.C. and M.M. Spectral pulse shaper section designed by D.N., N.F. and M.M, MPLC by N.F. and J.C., spatiotemporal holograms designed by A.V.K., M.M., N.F., J.C. and M.P., Optical system assembled by M.M, A.V.K., and N.F and aligned by A.V.K. Data analysis by A.V.K, M.M., N.F., M.P. and J.C. Manuscript written by M.M., A.V.K. and M.P. with input from all authors. M.M. conceived and supervised the project.

\subsection*{Ethics declarations}

The authors declare no competing interests.

\subsection*{Supplementary information}
\label{sec:supplementary}
Supplementary Notes 1–3, Figs. 1–3.

Supplementary Videos 1-7: The complete field animation for Fig.~\ref{fig:Toroidal Beams at arbitary orientation}


\end{document}